\documentclass[sigconf,authorversion]{acmart}

\usepackage{balance}
\usepackage{listings}
\usepackage{placeins}
\usepackage{enumitem}
\usepackage{microtype}
\usepackage{hyperref}

\newlist{researchquestions}{enumerate}{1}
\setlist[researchquestions]{label*=\textbf{RQ\arabic*.}}

\newlist{openquestions}{enumerate}{1}
\setlist[openquestions]{label*=\textbf{PSQ\arabic*.}}

\AtBeginDocument{%
  \providecommand\BibTeX{{%
    \normalfont B\kern-0.5em{\scshape i\kern-0.25em b}\kern-0.8em\TeX}}}

\copyrightyear{2025}
\acmYear{2025}
\setcopyright{acmlicensed}
\acmConference[ITiCSE 2025]{Proceedings of the 30th ACM Conference on Innovation and Technology in Computer Science Education V. 1}{June 27-July 2, 2025}{Nijmegen, Netherlands}
\acmBooktitle{Proceedings of the 30th ACM Conference on Innovation and Technology in Computer Science Education V. 1 (ITiCSE 2025), June 27-July 2, 2025, Nijmegen, Netherlands}
\acmDOI{10.1145/3724363.3729037}
\acmISBN{979-8-4007-1567-9/2025/06}

\begin{document}

\title[Broadening Participation through Physical Computing]{Broadening Participation through Physical Computing: Replicating Sensor-Based Programming Workshops for Rural Students in Sri Lanka}


\author{Poornima Meegammana}
\orcid{0009-0005-1419-5747}
\affiliation{%
  \institution{Faculty of Engineering and Design\\ University of Auckland}
  \city{Auckland}
  \country{New Zealand}}
\email{heshadharani.meegammana@auckland.ac.nz}

\author{Hussel Suriyaarachchi}
\orcid{0000-0002-8026-2523}
\affiliation{
\institution{Augmented Human Lab \\ National University of Singapore}
  \country{Singapore}
}
\email{hussel@ahlab.org}

\author{Paul Denny}
\orcid{0000-0002-5150-9806}
\affiliation{
  \institution{School of Computer Science \\ University of Auckland}
  \city{Auckland}
  \country{New Zealand}
}
\email{paul@cs.auckland.ac.nz}

\author{Suranga Nanayakkara}
\orcid{0000-0001-7441-5493}
\affiliation{
\institution{Augmented Human Lab \\National University of Singapore}
  \country{Singapore}
}
\email{suranga@ahlab.org}

\renewcommand{\shortauthors}{Poornima Meegammana et al.}


\begin{abstract}

In today's digital world, computing education offers critical opportunities, yet systemic inequities exclude under-represented communities, especially in rural, under-resourced regions. Early engagement is vital for building interest in computing careers and achieving equitable participation.  Recent work has shown that the use of sensor-enabled tools and block-based programming can improve engagement and self-efficacy for students from under-represented groups, but these findings lack replication in diverse, resource-constrained settings. This study addresses this gap by implementing sensor-based programming workshops with rural students in Sri Lanka.  Replicating methods from the literature, we conduct a between-group study (sensor vs. non-sensor) using Scratch and real-time environmental sensors.  We found that students in both groups reported significantly higher confidence in programming in Scratch after the workshop.  In addition, average changes in both self-efficacy and outcome expectancy were higher in the experimental (sensor) group than in the control (non-sensor) group, mirroring trends observed in the original study being replicated.  We also found that using the sensors helped to enhance creativity and inspired some students to express an interest in information and communications technology (ICT) careers, supporting the value of such hands-on activities in building programming confidence among under-represented groups.

\end{abstract}

\ccsdesc[300]{Social and professional topics~Computing education}

\keywords{Self-efficacy, Sensors, Under-represented, Block-based Programming, Physical Computing, Environmental Data}

\maketitle

\section{Introduction}
As computing and programming skills become essential in the modern world, ensuring that students from diverse backgrounds have equitable opportunities to succeed in the technology field is crucial. Many countries, including Australia, Italy, the United Kingdom, and Sri Lanka, have implemented national curricula to establish strong foundations in computing \cite{brown2014restart, duncan2015pilot,falkner2019international,lim2020information}. Despite these efforts, improving participation and outcomes among under-represented groups remains a global challenge \cite{lunn2021exploration,mcbroom2020understanding,newhall2014support}. Addressing these disparities is vital to creating a more inclusive technology sector.

One promising approach is the use of hands-on, engaging workshops, which provide opportunities for students to experience programming as both interesting and impactful \cite{freeman2014engaging, johnson2022underrepresented,allaireduquette2022gender}.  For example, Tshukudu et al. demonstrated that online programming workshops for students across several countries in Africa not only increased participants' confidence in programming but also highlighted the benefits of hands-on sessions guided by tutors \cite{tshukudu2022broadening}. Recent research has also shown that incorporating sensor-based tools in such workshops can make programming more relevant for under-represented groups \cite{husselsuriyaarachchi_2023_using}. Such tools offer a tangible and interactive way to learn, connecting coding concepts to real-world applications. However, to ensure that such findings are robust and generalised, there is a need to replicate this work in diverse contexts. Replication studies are a cornerstone of scientific inquiry, yet remain relatively rare in education \cite{perry17022022}.

In computing education, the lack of replication studies often hinders the field's progress, resulting in approaches that can be challenging to adopt in classroom practice across various disciplines \cite{ahadi2016replication,mcgill2019discovering,hao_2019_a}. Particularly in contexts that are less well studied, which may differ significantly from the well-resourced environments where most studies are conducted, replication is critical for advancing the field and addressing disparities in computing education. 

This motivated us to design a study to investigate the use of hands-on programming workshops for under-represented students from a rural community.  We replicated an approach described in prior work that utilised sensor-based toolkits. Following the methodology described by Suriyaarachchi et al.~\cite{husselsuriyaarachchi_2023_using}, we conducted workshops in Buttala, Sri Lanka. These workshops aimed to assess the impact of hands-on programming activities on self-efficacy and outcome expectancy of rural students aged 10--13. We seek to answer the following research questions:

\begin{researchquestions}
    \item How do measures of self-efficacy and outcome expectancy change among rural students in Sri Lanka after participating in a hands-on programming workshop, and does the use of a sensor-based toolkit influence these changes?
    \item What do students identify as the most enjoyable aspect of the workshop, and how does the sensor toolkit influence their decisions about program behaviour?
\end{researchquestions}

\section{Related Work}

\subsection{Sensors, Self-efficacy and Expectations}
Our research is a replication study of the work by Suriyaarachchi et al.~\cite{husselsuriyaarachchi_2023_using}, which investigated the use of sensor-based programming workshops for high school students from under-represented  groups in New Zealand, specifically M\=aori and Pasifika students. This original study demonstrated that sensor-based programming can enhance self-efficacy and outcome expectancy among under-represented students in New Zealand. The distinct educational landscape in Sri Lanka, including limited infrastructure, disparities in technology access, and the predominance of rural schools, provides a unique opportunity to explore the broader generalisability of the findings.  

In the study by Suriyaarachchi et al.~\cite{husselsuriyaarachchi_2023_using}, the authors explored a hands-on programming intervention using Scratch, a popular block-based programming environment, with two input modalities: standard keyboard/mouse controls and a sensor toolkit providing real-time environmental data. The study involved 49 participants divided into control and experimental groups. The experimental group used the sensor toolkit to create interactive programs responsive to environmental inputs, such as changes in humidity or light levels, while the non-sensor relied on traditional input methods. Pre- and post-workshop surveys measured self-efficacy and outcome expectancy using the Computer Science Attitudes Scale \cite{rachmatullah_2020_development}.

Their key findings showed that participation in the workshops improved self-efficacy for all students, regardless of input modality. However, outcome expectancy improved significantly only for students in the sensor group. Additionally, qualitative feedback revealed that students enjoyed the workshop, with sensor-based activities sparking interest and creativity. This underscores the potential of sensor-based programming to engage under-represented groups in computing education by making programming more tangible and interactive. It also suggests a need for culturally and contextually relevant interventions to bridge equity gaps in computing education. The current replication builds on these findings by exploring the generalisability of sensor-based programming in a different cultural and resource-constrained context.

\subsection{Sensor use in programming education}
Programming education increasingly uses sensor technology to connect coding concepts with real-world applications. Sensors are integrated in various settings, including classrooms and STEM camps. Most commonly used sensors, such as light, sound, and distance sensors (e.g.KIBO robots), are effective tools for teaching coding concepts \cite{bers2022coding,sullivan2018dancing}. Environmental sensors, which measure temperature, humidity, CO$_2$, and soil moisture sensors, as seen with products like Micro:bit and Gator:bit, introduce students to environmental monitoring and data analysis \cite{chakarov2020opening}. These tools enable students to connect programming with physical phenomena, creating interdisciplinary learning opportunities. Programming environments also play a critical role in integrating sensors into educational settings. Visual programming platforms like MakeCode simplify sensor programming for younger learners, while tangible tools such as KIBO and FYO robots, provide engaging hands-on experiences \cite{chakarov2020opening,bers2022coding}. For older students, text-based programming environments like the Arduino IDE offer more advanced, syntax-driven coding opportunities suitable for varying skill levels \cite{kafai2019stitching}. 

Research also highlights the positive impact of integrating sensors into programming education. Sensors enhance interactivity, sustaining student engagement over extended periods \cite{bers2022coding}. They also promote creativity, confidence, problem-solving, and data analysis skills \cite{bicer2018cracking,bondaryk2021probeware,kafai2019stitching,lee2020effectiveness}. Additionally, sensors help learners grasp complex concepts, such as data transmission and environmental monitoring \cite{bicer2018cracking}.
However, challenges persist. For example, rotation sensors can sometimes create confusion, underscoring the importance of clear instructional design \cite{caceres2018tangible}. Despite these challenges, educational programs that incorporate sensors consistently report high levels of student motivation and engagement, largely due to the tangible outcomes that students achieve through coding.

\subsection{Under-represented groups in computing}
The under-representation of certain groups in STEM disciplines, particularly in computing, remains a significant and ongoing challenge at both the secondary and tertiary education levels \cite{Guzdial2012,margolis2012beyond}. Research highlights that women, ethnic minorities, underprivileged groups, and individuals with disabilities are disproportionately under-represented in these fields. Contributing factors include prevalent societal stereotypes about gender and race, a lack of relatable role models, and limited access to quality educational resources. Warschauer et al.~\cite{warschauer2004technology} emphasise that issues related to the digital divide can diminish confidence and lower expectations of success among underprivileged students in computing fields. This underscores the critical impact of equitable access to digital and educational resources on participation and achievement in STEM. In the context of Sri Lanka, disparities in computing education are particularly evident between urban and rural areas. Rural schools frequently lack the necessary infrastructure, qualified educators, and access to high-quality learning resources, placing them at a significant disadvantage compared to urban institutions \cite{pratheepan2019study}. Mentorship has emerged as one effective intervention to address underrepresentation by fostering scientific identity and creating career pathways for marginalised groups \cite{atkins2020}. Ongoing work is essential to continue addressing systemic inequities and promoting greater diversity in STEM fields.

\section{Methodology}
In our study, we employed a between-group design. The independent variable was the integration of a sensor kit-based programming lesson, which divided participants into two groups: a non-sensor group (control group; $n = 62$) and a sensor group (experiment group; $n = 60$). Data collection involved administering pre-study questionnaires to establish baseline measurements and post-study questionnaires to evaluate changes. The questionnaires assessed participants’ computer programming self-efficacy, outcome expectancy, and confidence levels.

\subsection{Participants}
We partnered with the Shilpa Sayura Foundation to conduct this study at one of their training locations in Buttala, Sri Lanka. Buttala is a rural town in the Monaragala District, where agriculture serves as the primary livelihood for most residents. The study was designed as a workshop aimed at increasing confidence among rural students aged 10-13 to pursue careers in technology. 
This community represents an under-represented group in technology, making it appropriate for our intervention. We organised the workshop in two sessions with a total of 122 participants ($n = 122$). The average age of the participants was 11.4 years, with a standard deviation of 1.1. Notably, 69\% of participants did not have access to a computer or laptop at home. Additionally, 53\%  reported using a computer ``not very often'' or  ``never'' while only 11\% stated that they used a computer ``every day'' or ``most days''. Furthermore, 90\% of the students mentioned that they seen or used Scratch in a group or class outside of school.

\subsection{Programming environment  and sensors}
In this study, we used the Scratch programming environment due to its accessibility and ease of use, particularly for children and adolescents learning to program \cite{monroyhernndez_2008_featureempowering}. 
The sensor group used an extension of the Scratch programming interface developed by Kiwrious\footnote{\url{https://kiwrious.com}}, which is freely accessible online and which supports the connection of low-cost environmental sensors via USB \cite{suriyaarachchi2022scratch, suriyaarachchi2022primary}.

\begin{figure}[t]
    \centering
    \includegraphics[width=\linewidth]{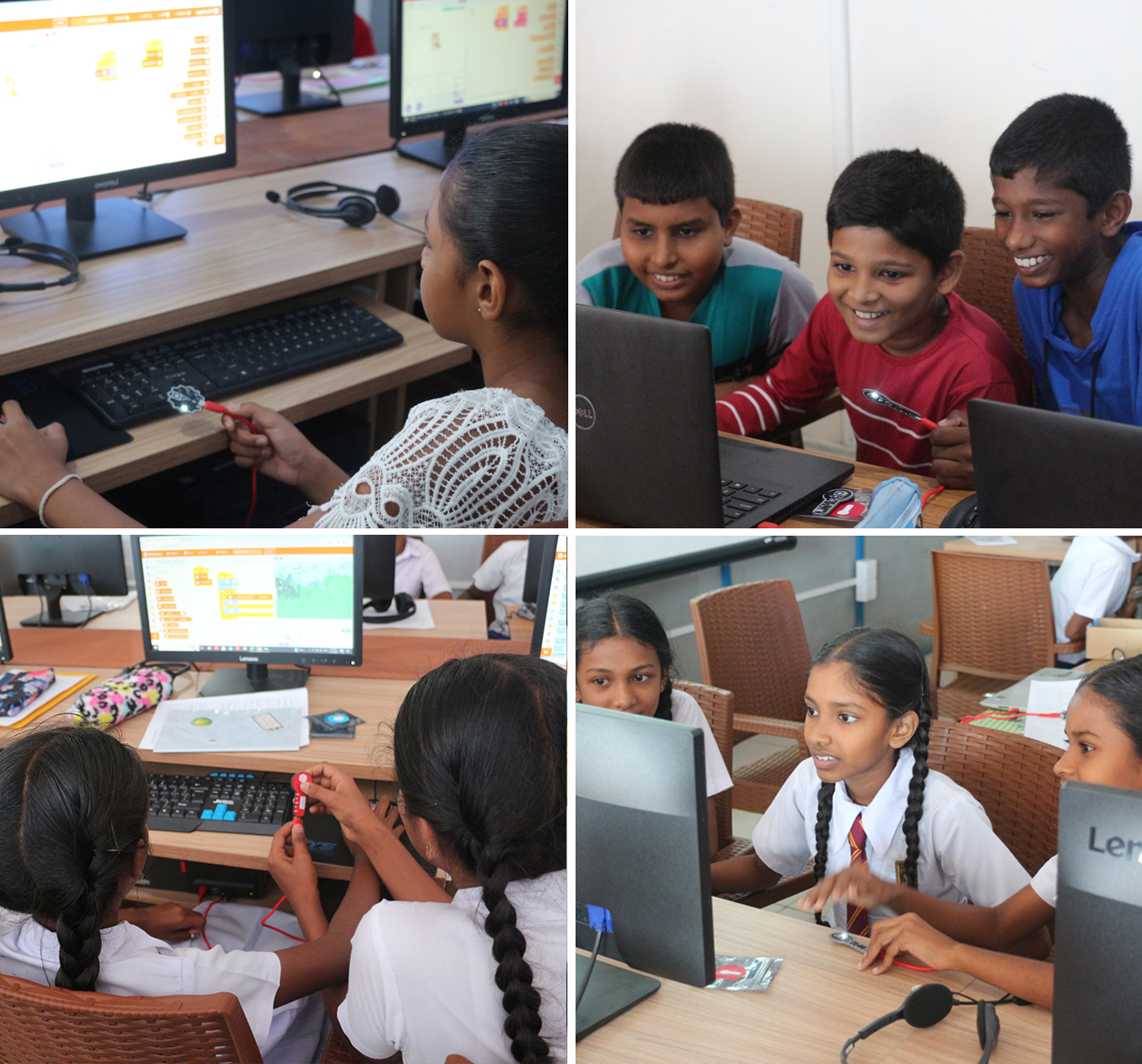}
    \caption {
Students in the sensor group using the sensor toolkit during a hands-on programming workshop. Permission to use these images has been granted.}
    \label{fig:session}
    \vspace{-10pt}
\end{figure}

The sensor toolkit we used includes devices such as humidity sensors, surface temperature sensors and visible light sensors, all of which provide real-time input to executing programs. These sensors plug into a computer using a USB port and are automatically detected as a plug-and-play device. Figure \ref{fig:session} depicts students actively working with the sensor toolkit during the workshop. 

\subsection{Study Procedure}
Students were randomly assigned to either the experiment (sensor) or control (non-sensor) group and a separate workshop was run for each group. 
Each of these two sessions lasted 40 minutes and included an introduction to programming with Scratch (10 minutes; see Section \ref{MethodIntro}), open-ended coding activity (20 minutes; see Section \ref{MethodActivity}), and data collection (10 minutes; see Section \ref{MethodData}). In terms of practically running each session, the students worked in small groups of three on the programming tasks. 
Both sessions were conducted in Sinhala, the local language, which was also used in the questionnaires provided to the students.

In the experiment (sensor) group, each small group of three students were provided with a sensor kit containing three sensors selected from temperature, conductance, humidity, and visible light sensors. We demonstrated how to connect the sensors to the computer, add sensor-specific blocks to their Scratch programs and use sensor-based data to control the sprite. In the control (non-sensor) group, students also worked in small groups of three and were shown how to use the keyboard and mouse as input to control the sprite. Both the control and experimental groups worked with identical interfaces, sprites, and code blocks, with the exception of the sensor-specific code blocks, which were only available to the sensor group.

\subsubsection{Introduction to Programming} \label{MethodIntro}
\begin{figure}[t]
    \centering
    \includegraphics[width=\linewidth]{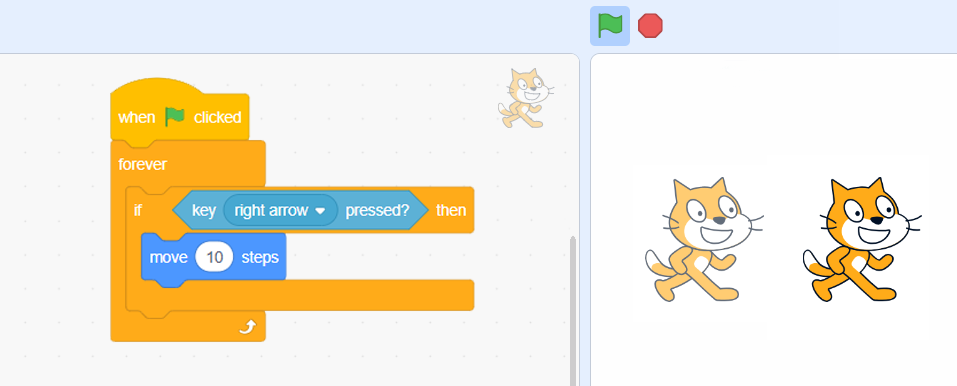}
    \caption{ The Sample Scratch program for the non-sensor group. The sprite is controlled by keyboard input and it moves to the right when the right arrow key is pressed
    }
    \label{fig:Scratch}
\end{figure}

\begin{figure}[t]
    \centering
    \includegraphics[width=\linewidth]{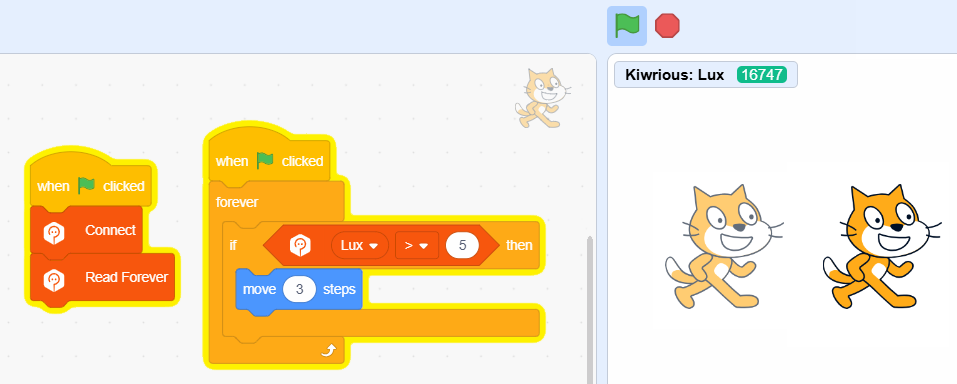}
    \caption{ The Sample Scratch program for the sensor group. The sprite is controlled by the visible light sensor and moves to the right when light intensity is greater than 5.}
    \label{fig:sensor}
\end{figure}

Participants were introduced to Scratch through a basic overview of connecting blocks and a sample program that demonstrates the core components and functions of Scratch.  The only difference between the sample programs for the control and sensor groups was in the sprite control method. The non-sensor group (control group) used conventional key presses (Figure \ref{fig:Scratch}) for sprite control, while the sensor group (experiment group) used a sensor block (Figure \ref{fig:sensor}) to trigger sprite movement based on changes in visible light changes. Despite differing input modalities, the sprite's behaviour was consistent across both groups.

\subsubsection{Programming Activity}  \label{MethodActivity}
After the introduction, the students spent 20 minutes exploring and creating their own programs. To ensure a valid comparison for our evaluation, we instructed students to include at least one element controlled by an input specific to their assigned study group. Keyboard/mouse for the non-sensor group and sensors for the sensor group. Four instructors were present during the workshop to assist the students as needed.

\subsubsection{Data Collection}  \label{MethodData}
For data collection, we designed pre- and post-questionnaires containing Likert-scale items in the local language. The questionnaires were printed to make it easy for students to fill them out. The questionnaire included demographic questions along with three self-efficacy questions (I1–I3), six outcome expectancy questions (I4–I9), and one question about confidence in programming with Scratch (Q1). Items I1–I9 were adapted from the Computer Science Attitudes Scale \cite{rachmatullah_2020_development}. Table \ref{tbl:questions} lists the English translations of each question (which originally appeared on the questionnaires in Sinhala). 

\begin{table}[b]
    \vspace{-10pt}
    \caption{Questionnaire items (presented to participants in Sinhala) translated to English}
    \vspace{-10pt}
    \centering
    \includegraphics[width=0.9\linewidth]{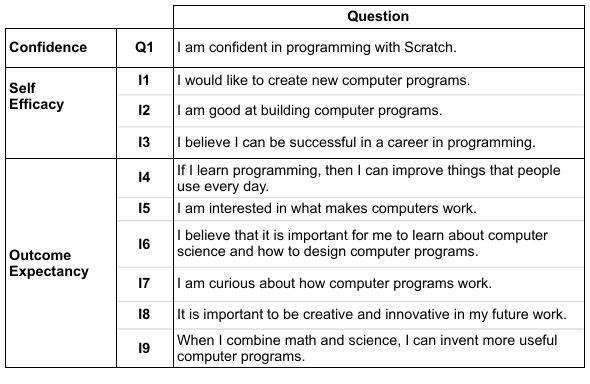}
    \label{tbl:questions}
\end{table}

The pre-questionnaire was administered prior to the ``Introduction to Scratch'' session, and the post-questionnaire was given immediately after the students completed their programming activity. All responses were recorded on a 5-point Likert-scale, ranging from 1 (strongly disagree) to 5 (strongly agree). 

The post-study questionnaire also included two open-ended questions:
\begin{openquestions}
    \item Tell us what you enjoyed most about today’s activity.
    \item Tell us a little about your Scratch project. How did you control your character(s)? Why did you program them that way?
\end{openquestions}
We encouraged students to answer all questions honestly and to the best of their ability, assuring them that there were no right or wrong answers.

\subsubsection{Data Analysis}
We collected pre- and post-questionnaire data from a total of 122 participants ($n = 62$ non-sensor; $n = 60$ sensor). However, 5 data sets were incomplete and were subsequently excluded from the analysis, resulting in a data set comprising 2,340 individual data points. For each participant, we determined self-efficacy and outcome expectancy by averaging their responses to specific questionnaire items. Confidence was measured using responses to Q1, self-efficacy was calculated from items I1–I3, and outcome expectancy was derived from items I4–I9. 
We followed the recommendations of Rachmatullah et al.~\cite{rachmatullah_2020_development}, whose survey instrument we used to compute these metrics.
This approach allowed us to generate paired pre- and post-measurements for each construct, facilitating a direct comparison of changes across the study. That is, for each group (control and experimental), we compute the change in confidence, self-efficacy and outcome expectancy (before and after the workshop) for each student in that group. 

We used the Wilcoxon signed-rank test, a nonparametric equivalent of the paired samples t-test, to analyse the data \cite{c4091bd3-d888-3152-8886-c284bf66a93a}.
A Bonferroni correction was applied to account for the multiple comparisons performed on the questionnaire data and to ensure appropriate adjustment of the significance levels for our statistical tests.
Since we conducted a Wilcoxon signed-rank test for each construct (confidence, self-efficacy, and outcome expectancy), the standard significance threshold ($p = 0.05$) was divided by 3, resulting in an adjusted $p$-value of $0.017$.
In addition to our quantitative analysis, we conducted a deductive thematic analysis of responses to PSQ1 and PSQ2 to explore what students enjoyed about the session. 
As we sought to replicate the work of Suriyaarachchi et al.~\cite{husselsuriyaarachchi_2023_using}, we followed a deductive coding process, categorising responses to the open-ended questions using the same three themes identified in their study.
As qualitative responses were in Sinhala, a native-speaking research assistant proficient in English translated the data, which was then reviewed by the researcher, also a native Sinhala speaker, to ensure accuracy before thematic coding.

\section{RESULTS \& DISCUSSION}
This section presents the results of the workshop data analysis, organised into three parts. First, we report findings related to confidence in programming (Q1), self-efficacy (I1–I3) and outcome expectancy (I4–I9). 
Next, we discuss qualitative insights based on students’ open-response feedback, focusing on what they enjoyed during the sessions and drawn from the students' descriptions of their projects.

\subsection{Student Attitudes and Perceptions}
A summary of our statistical analysis of the changes in confidence, self-efficacy, and outcome expectancy among students in the control (non-sensor) and experimental (sensor) groups is presented in Table \ref{tbl:stat}.
We observed a statistically significant improvement ($p < .0001$) in the programming confidence reported by students in both the non-sensor and sensor groups. Responses to questionnaire item Q1 from the pre-study survey indicated a generally neutral sentiment about programming confidence. 
As expected, students felt more confident programming in Scratch by the end of the workshop, with the sensor group showing a larger increase in the mean difference ($\mu_{\text{diff}}$=1.11) compared to the non-sensor group ($\mu_{\text{diff}}$=0.77). 

Similarly, the overall average changes in self-efficacy and outcome expectancy were higher in the experimental (sensor) group than in the control (non-sensor) group.
We found no significant difference between the non-sensor group's pre-and post-test self-efficacy scores after applying the Bonferroni correction ($p=.043$).
Meanwhile, students in the experimental group demonstrated a significant improvement in their perception of self-efficacy ($p=.007$).
For outcome expectancy, neither group exhibited a significant difference between their pre-and post-test scores. However, students in the non-sensor group experienced a slight decrease in outcome expectancy, whereas those in the experimental group were essentially unchanged.

These findings agree with the results reported by 
Suriyaarachchi et al.~\cite{husselsuriyaarachchi_2023_using} for their study conducted in the New Zealand context. 
This suggests that the use of programming workshops, especially involving sensor-based inputs, may be an effective intervention for under-represented communities, particularly in enhancing confidence and self-efficacy.
We do not replicate the findings related to outcome expectancy (where Suriyaarachchi et al.~\cite{husselsuriyaarachchi_2023_using} observed a difference between control and experimental groups). 
One possible explanation for this could be the initially high pre-study self-report scores for this measure in both the control ($\mu_{pre}=3.99/5$) and experimental groups ($\mu_{pre}=4.02/5$). Since most participants in our study had been engaged in a STEM program for some time, they may have already developed a positive perception of the outcomes they could achieve through learning technology-related skills.

\begin{table}[b]
    \vspace{-10pt}
    \caption{Analysis of the pre and post results within each group using a Wilcoxon Signed Rank test. $\mu_\textit{diff}=\textit{post-pre}$, *= significant with Bonferroni-corrected $p<0.017$.}
    \vspace{-10pt}
    \centering
    \includegraphics[width=.9\linewidth]{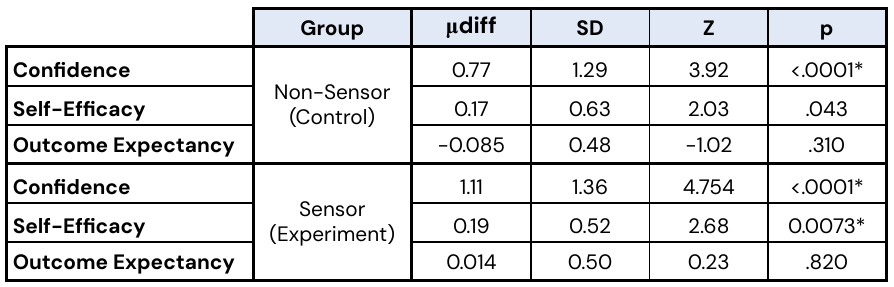}
    \label{tbl:stat}
\end{table}

\begin{figure}[t]
    \centering
    \includegraphics[width=0.8\linewidth]{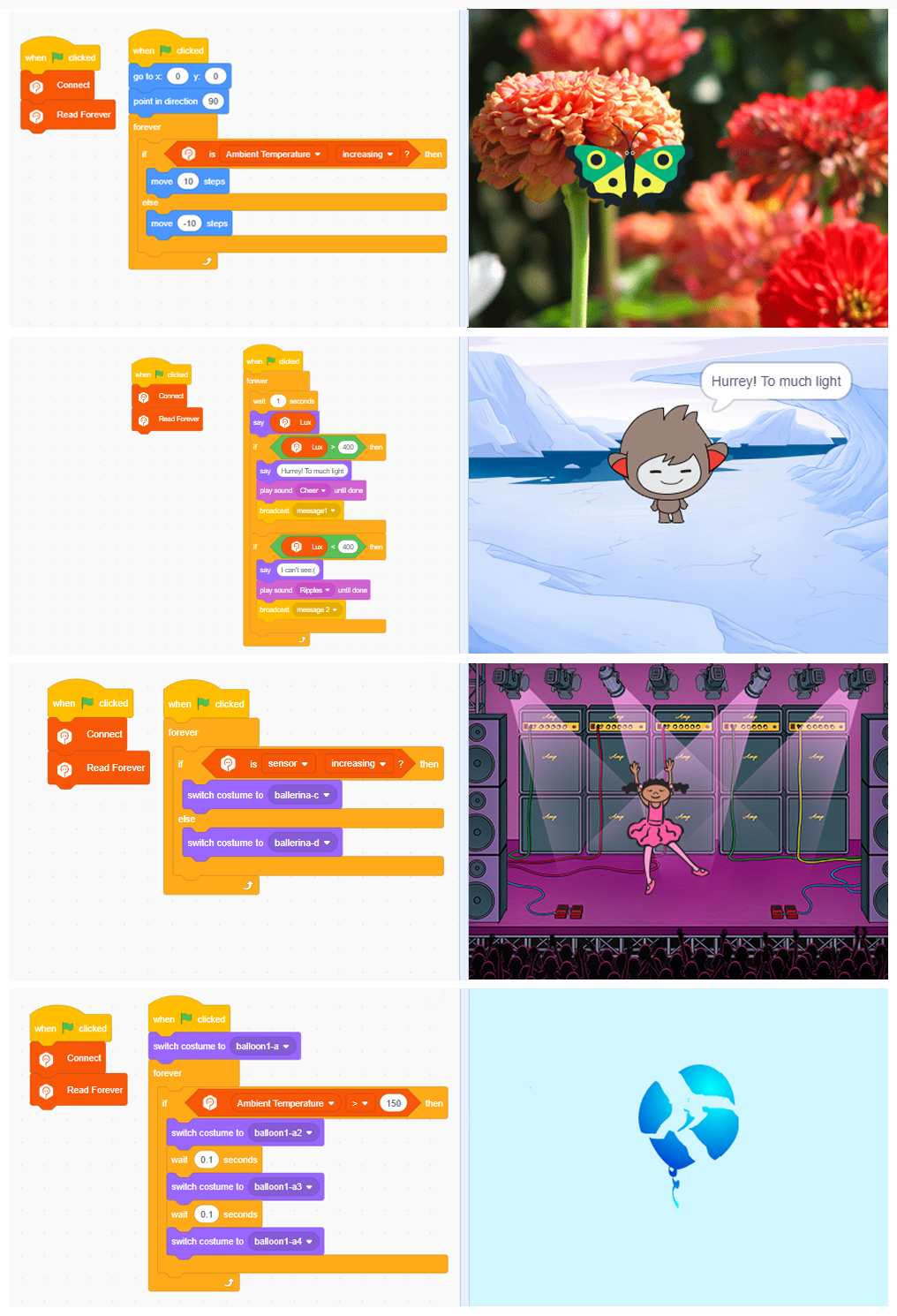}
    \caption{ A Sample of Student Projects from the Workshop.}
    \label{fig:Studentproj}
\end{figure}

\subsection{The Most Enjoyable Aspects of the Activity}
This section reports the qualitative results of open-response feedback from PSQ1 and PSQ2. Using a deductive approach, we applied the three themes identified by Suriyaarachchi et al.~\cite{husselsuriyaarachchi_2023_using} around opportunities for `Learning', `Creating', and `Interacting' -- to tag each comment. These themes were used to analyse responses from both the non-sensor (control) group and the sensor (experimental) group.


\subsubsection{Learning}
A significant proportion of students in the non-sensor group (37\%) identified learning programming concepts as the most enjoyable aspect of the workshop. Many responses praised the simplicity and accessibility of Scratch as an introductory tool. For instance, one student shared, \textit{``I thought programming was difficult, but it was very easy. I like to do programming.''}

In the sensor group, 27\% of students mentioned learning programming as a highlight, though this percentage was lower compared to the non-sensor group. Instead, 60\% of the students in the sensor group expressed excitement about integrating sensors and 38\% enjoyed incorporating environmental data into programming activities. For example, one student commented, \textit{``I liked the activity we did using a light sensor. Today, we learned something we did not know before,''} while another said, \textit{``Using `if-then' to say `Hurry! Too much light' when the light level is above 400 and `I can’t see' when it's below 400.''} These responses suggest that the sensor toolkit added a novel dimension to programming, shifting the focus towards interactive and sensor-driven activities. Students in the sensor group also noted the novelty of the experience, with 12\% describing it as the most enjoyable aspect, compared to 5\% in the non-sensor group. One student stated, \textit{``I liked that we used something we had never seen or heard of,''} and another reflected, \textit{``This is the first time I made an application. It’s an unforgettable experience for me.''} 

Additionally, 7\% of students in the sensor group explicitly mentioned an interest in pursuing ICT as a future field of study or career, showcasing the workshop’s inspirational impact. For instance, one student remarked, \textit{``I liked the sensor we used. I would like to be a Scratch teacher in the future''} (i.e., teach programming), while another shared, \textit{``ICT is a beautiful and surprising subject. I would like to do more like this.''}

\subsubsection{Creating}
The creative aspects of the task were 
frequently highlighted, with 56\% of non-sensor and 75\% of sensor students identifying them as the most enjoyable aspect. Participants in the non-sensor group particularly appreciated activities such as moving sprites, drawing shapes, and programming sprites to perform specific actions. For example, one student noted, \textit{``I liked drawing shapes, jumping, and running,''} referring to sprite actions. Responses to PSQ2 revealed that 61\% of the non-sensor group incorporated creative elements, including character creation (38\%) and storytelling (10\%).  For example, one student described, \textit{``My program was to save an insect from a hungry fox.''}

In the sensor group, creative activities often involved sensor integration. For instance, one student noted, \textit{``I enjoyed using the sensor and the way the cat ran. I like making projects,''} and another shared, \textit{``I liked using the light sensor to make a dialogue''}. From PSQ2, 71\% of responses emphasised creative elements, including background creation (18\%) and character development (45\%). For example, one student explained, \textit{``We made an underwater scene with fish and controlled the fish with the temperature sensor.''}.

\subsubsection{Interacting}
Interactive aspects of the workshop were enjoyed by 48\% of students in the non-sensor group and 50\% in the sensor group. Non-sensor group students highlighted their enjoyment of using Scratch blocks, keyboard inputs, and mouse inputs to control sprites. For example, one student said, \textit{``I liked that I was able to make an insect go up, down, left, and right.''}
In the sensor group, students enjoyed observing how sensor inputs affected sprite behaviour. For instance, one student remarked, \textit{``My sprite moved when the sensor detected a temperature increase,''} while another noted, \textit{``The cat running slow with low light and running fast when there is more light.''} Sprite actions also played a significant role in how students interacted with the program. In the non-sensor group, 40\% of  PSQ2 responses highlighted sprite actions such as moving (37\%) and rotating (4\%) characters. In comparison, 58\% of sensor group participants described sensor-driven actions, such as sprites moving in response to environmental data. For example, one student explained, \textit{``The balloon bursts when the light increases beyond 150.''}

The results reveal that students found distinct elements enjoyable based on whether they participated in the non-sensor (control) group or the sensor (experimental) group. Students in the non-sensor group primarily enjoyed acquiring programming skills, exploring Scratch, and creative freedom. In contrast, the sensor group reported enjoyment in activities that involved the integration of sensors. This suggests that the sensor toolkit redirected students' focus towards interactive and creative applications of technology, enhancing their programming experience. The overarching themes of learning, creating, and interacting were evident in both groups, aligning with the findings of Suriyaarachchi et al. ~\cite{husselsuriyaarachchi_2023_using}.

Several students in the sensor group expressed enthusiasm for exploring ICT as a potential field of study or career.  Use of the sensor toolkit for the programming activity appeared to enhance both their immediate engagement and longer-term aspirations in STEM.

%

\subsection{Limitations and Future Work}
Our aim was to replicate prior work from the computing education literature. 
Although we observed increases in confidence (significant in both groups) and self-efficacy (significant in the experimental group), mirroring previous results, we remain cautious in overstating these findings. The sample size ($n=122$) from a single rural community in Sri Lanka limits generalisability, highlighting the need for broader, multi-context studies. Participants’ limited prior programming exposure and reliance on self-reported questionnaires may have influenced findings, suggesting future work should account for technological familiarity and include more objective measures. Incorporating additional measures like physiological data, task completion time, or follow-up interviews could provide more objective and deeper insights into the long-term effects on students' ICT interests and aspirations. The workshops were limited to 40-minute sessions, which may have constrained students’ ability to fully internalise the concepts. Extending the duration or conducting longitudinal studies may provide a more comprehensive assessment of sustained impact. Furthermore, the absence of discussions on ICT career pathways or relatable role models represent a missed opportunity to inspire students. Relatable success stories are particularly impactful for under-represented groups, as they help students envision potential futures in technology fields \cite{zirkel2002there}.

\section{Conclusion}
In this paper, we report on a sensor-based programming workshop conducted among rural students in Sri Lanka, designed to replicate prior findings from research in a more well-resourced context. We investigated how programming confidence, self-efficacy, and outcome expectancy changed during the workshop and examined the influence of sensor-enabled tools on these outcomes.
Our results indicate that the workshop improved programming confidence for all students, with the sensor group showing greater gains across all measures compared to the non-sensor group.
Notably, self-efficacy improved significantly among students who used the sensor toolkit to access environmental data during their programming tasks.
Qualitative feedback revealed that integrating sensors into programming tasks enhanced creativity and interactivity, inspiring some students to consider ICT careers. These findings support the potential of sensor-based programming to promote inclusivity in computing education for under-represented groups, even in resource-constrained settings. By making programming tangible and interactive, sensor-enabled activities can foster greater engagement and help bridge equity gaps in STEM fields.

\balance



\bibliographystyle{ACM-Reference-Format}
\bibliography{sample-base}

\end{document}